\documentclass[]{spie}  

\usepackage{amsmath,amsfonts,amssymb}
\usepackage{graphicx}
\usepackage[colorlinks=true, allcolors=blue]{hyperref}

\usepackage[numbers]{natbib}
\usepackage{notoccite}
\usepackage[dvipsnames]{xcolor}

\usepackage{soul}
\usepackage{booktabs}
\usepackage{textgreek}
\usepackage{threeparttable}

\title{Weakly Supervised Attention Model for RV Strain Classification from volumetric CTPA Scans}

\author[a]{Noa Cahan}
\author[b]{Edith M. Marom}
\author[b]{Shelly Soffer}
\author[b]{Yiftach Barash}
\author[b]{Eli Konen}
\author[b]{Eyal Klang}
\author[a]{Hayit Greenspan}

\affil[a]{Department of Biomedical Engineering, Tel-Aviv University, Tel-Aviv, Israel}
\affil[b]{Department of Diagnostic Imaging, Sheba Medical Center, Ramat Gan, Israel affiliated with the Tel Aviv University, Tel Aviv, Israel}

\begin{document} 
\maketitle


\begin{abstract}

Pulmonary embolism (PE) is a life-threatening condition, often without warning signs or symptoms. Early diagnosis and accurate risk stratification are critical for decreasing mortality rates. High-risk PE relies on the presence of right ventricular (RV) dysfunction resulting from acute pressure overload. PE severity classification and specifically, high-risk PE diagnosis are crucial for appropriate therapy. Computed tomography pulmonary angiography (CTPA) is the golden standard in the diagnostic workup of suspected PE. Therefore, it can link between diagnosis and risk stratification strategies.

In this work, we address the problem of RV strain classification from 3D CTPA scans. We retrieved data of consecutive patients who underwent CTPA and were diagnosed with PE. We extracted a single binary label of ``RV strain biomarker" from the CTPA scan report. This label was used as a weak label for classification. Our solution applies a 3D DenseNet network architecture, further improved by integrating residual attention blocks into the network’s layers. 

This model achieved an area under the receiver operating characteristic curve (AUC) of 0.88 for classifying RV strain. For Youden's index, the model showed a sensitivity of 87\% and specificity of 83.7\%.  Our solution outperforms state-of-the-art 3D CNN networks. 
The proposed design allows for a fully automated network that can be trained easily in an end-to-end manner without requiring computationally intensive and time-consuming preprocessing or strenuous labeling of the data. This current solution demonstrates that a small dataset of readily available unmarked CTPAs can be used for effective RV strain classification. To our knowledge, this is the first work that attempts to solve the problem of RV strain classification from CTPA scans and this is the first work where medical images are used in such an architecture. Our generalized self-attention blocks can be incorporated into various existing classification architectures making this a general methodology that can be applied to 3D medical datasets.
\end{abstract}


\section{INTRODUCTION}


PE refers to obstruction of pulmonary arteries by blood clots. Among all cardiovascular diseases, PE is the third most common cause of death, after coronary heart disease and stroke. It accounts for approximately 100,000 deaths per year in the United States alone \citep{pe_mortality}. The clinical presentation of PE is variable and often nonspecific, making the diagnosis challenging \citep{UpToDatePe}. Thus, rapid diagnosis and accurate risk stratification are of paramount importance.
High-risk PE is caused by RV dysfunction from acute pressure overload. It is predictive of in-hospital mortality and identifying the need for more aggressive therapy. RV enlargement can be detected via CTPA, which also provides the diagnosis of PE. Therefore, CTPA can link between diagnosis and risk stratification \citep{cite-key-pe-rv, 0002-870390155-8}.


A major limitation of imaging in PE diagnosis and in assessing treatment response is the lack of robust quantitative assessment. Measurements are currently done manually, which is laborious and involves inter-reader variation. 
Computerized analysis can support the clinical workflow by enabling quantitative, reproducible evaluation.

Deep learning applications have shown success in various medical imaging tasks such as Chest x-ray classification \citep{DBLP:journals/corr/abs-1711-05225}, pancreas segmentation \citep{oktay2018attention} and anomaly segmentation in brain MRI \citep{Baur_2019}. Our task of automated risk stratification of PE still poses a challenge for researchers. The variability PE appearance, and lack of public datasets, make PE distinction difficult \citep{10.1038, masoudi2017dataset, cad_pe}. 
Most existing works on PE focus on the detection task. Computer-aided detection (CAD) systems have been suggested to detect the presence of PE. Most used convolutional neural networks (CNNs) for classifying PE candidates that are first extracted from an entire CTPA volume based on voxel-level features using traditional feature engineering methodologies. While partially successful, a few major limitations include the need for manual feature engineering, complex preprocessing, a high number of false positives, and an uncertainty degree in generalization \citep{10.1007/978-3-540-73273-0_52, cta, 5540279, 2012SPIE.8315E..2UW, multi_planar_pe}. A more recent solution from Yang et al.\citep{two_stage_pe} presented a cascaded two-stage CNN achieving a high sensitivity of 75\%. However, the model requires the division of the CTPA into small cubes rather than processing the entire CTPA scan. This approach still involves some preprocessing and suffers from a lack of an ``end-to end" solution required for a clinical application. PENet \citep{PENet} is a 3D CNN that aims to detect the PE in a series of slices from a CTPA study and currently achieves state-of-the-art performance for PE detection. It processes each slice using a 2D CNN and then uses a 3D CNN to aggregate information from many consecutive slices. This solution requires the labeling of each of the CTPA scan slices individually for the training process. The latest publication from Shi et al.\citep{shi2020automatic}, improves AUC accuracy compared to PENet by adding supervised attention maps that use pixel-level annotations for a small set of data.
Other works monitor RV motion or enlargement in MRI scans \citep{inproceedings, RV-MRI, bernardino2020volumetric}, or predict pulmonary hypertension in CTPA scans \citep{10.1117/12.2512469}. In all the papers presented above, the methods are either fully supervised where the datasets contain full segmentation maps or are labeled on a 2D slice-level and require heavy preprocessing of the data.

A frequent problem when applying deep learning methods to medical images is the lack of labeled data. Manual labeling of images is an expensive and time-consuming process. This issue motivates approaches beyond traditional supervised learning by incorporating other data and/or labels that might be available. These approaches include semi-supervised, weakly-supervised, multiple instances, and transfer learning \citep{DBLP:journals/corr/abs-1804-06353}. In the current study, instead of relying on manually produced pixel-level annotations, a single patient-level label is extracted from radiology reports. This distinguishes our design as a fully automated network that can be trained in an end-to-end manner without requiring preprocessing or exhausting labeling of the data.

We developed a weakly supervised deep learning algorithm, with an emphasis on a novel attention mechanism, to automatically classify RV strain from 3D CTPA scans. 
We demonstrate the implementation of attention modules in various 3D residual and densely connected network architectures baselines. We compare our model to state-of-the-art 3D CNN classification models. The results show that attention consistently improves prediction accuracy across different models. Also, our model performance surpasses all compared classification networks. The main contributions of this paper can be summarized in the following three aspects: \begin{enumerate}
\item \textbf{RV strain classification from volumetric CTPA scans.} To the best of our knowledge, there is no prior deep learning-based solution for fully automated classification of RV strain (or PE severity assessment) using contrast-enhanced chest CT scans.
\item \textbf{Novel 3D DenseNet with residual attention blocks.} Our method focuses on integrating deep residual attention blocks into a 3D dense connection block-based model.
We present two models that share the same backbone but differ in the attention-block integration. 

\item \textbf{Weakly supervised end-to-end solution.} 
The dataset we use is annotated by a single scan-level label for the whole 3D scan, with no additional markings or even any slice-level annotation, making this method exceptionally weakly supervised. The proposed design allows for a fully automated network in which the 3D volumetric scans are fed to the network as a whole with only minimal preprocessing, thereby, can be trained easily in an end-to-end manner.


\end{enumerate}

\section{METHODOLOGY}
We propose and compare two possible 3D DenseNet network architectures that we combine with residual attention modules for a novel solution to the analysis of RV strain from 3D CTPA images.

\subsection{3D DenseNet with Residual Attention Blocks Architecture}
\label{sec:architectures} 

The network’s backbone is a 3D DenseNet architecture \citep{8578783}, further improved by residual attention blocks. The 3D DenseNet Attention model is inspired by the residual attention network for image classification \citep{8100166} and hand gesture recognition videos \citep{20.500.11850/365762}. In these works, attention modules were integrated into the deep residual network’s layers to generate attention-aware features. Our solution generalizes this design from 2D to 3D and uses a baseline of densely connected blocks rather than residual ones. We propose soft attention, image-grid based gating that allows attention coefficients to center on local regions. Four 3D residual attention blocks are incorporated into the network. Each attention block captures different types of features extracted from a different layer in the baseline network.

\begin{figure}[ht]
   \begin{center}
   \begin{tabular}{c} 
   \includegraphics[height=6cm]{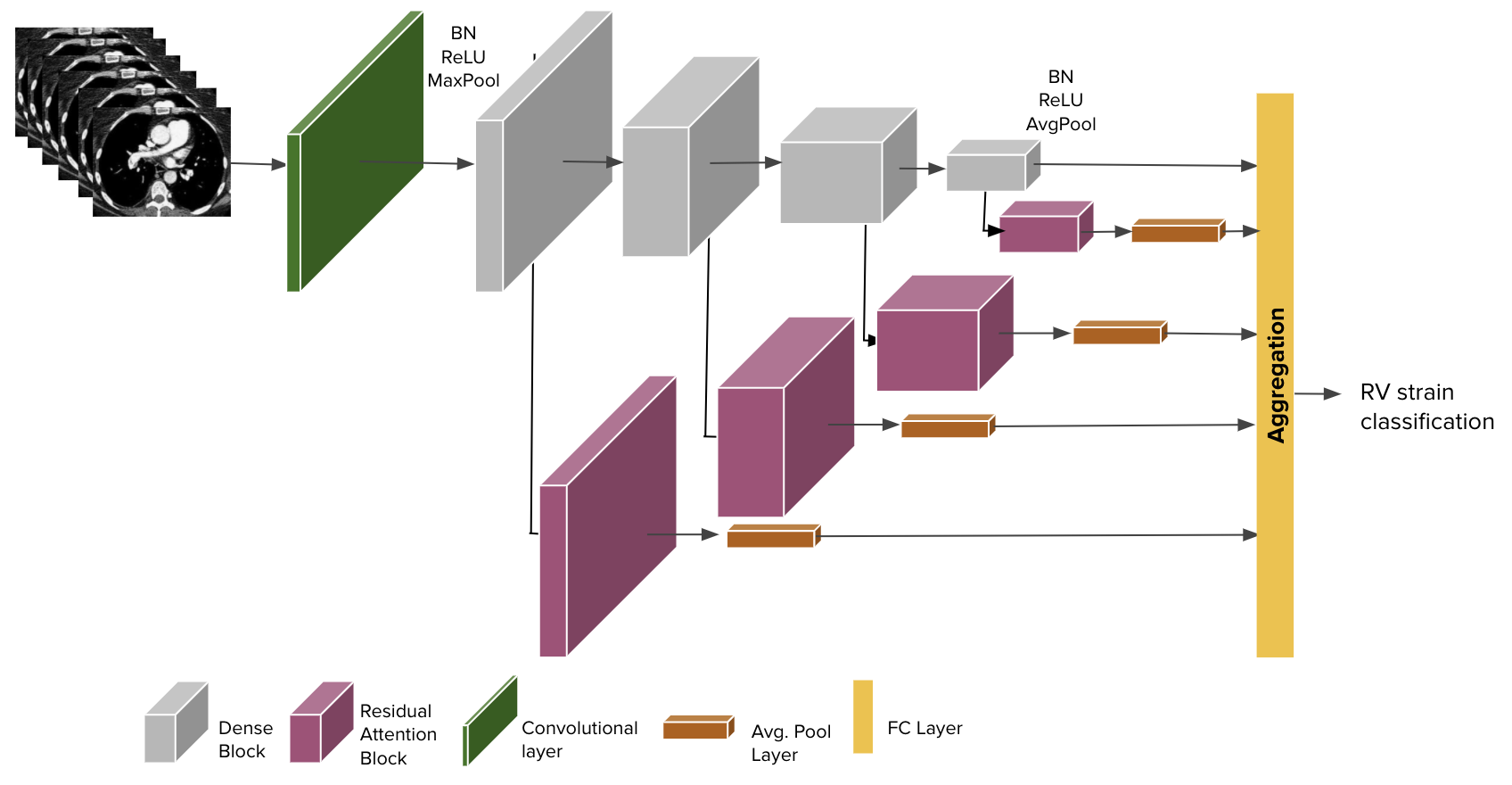}
   \end{tabular}
   \end{center}
   \caption[example] 
   { \label{fig:AttentionBlockOption2} 
MLANet: 3D DenseNet Multi-layer attention model block diagram}
   \end{figure}

We present two architectures constructed from the same baseline but differ in how the attention blocks are integrated. In the first architecture -- \textit{Multi-Layer Attention Network} (MLANet), we integrate the blocks in a more common practice as first introduced in \citep{jetley2018learn}. This is a multi-layer feature extraction approach. The baseline network is used as a discriminant feature extractor, and several attention blocks are applied on extracted features from different layers throughout the network to create attended feature vectors at different scales. The attended feature vectors are flattened via channel-wise global average pooling and combined for the final prediction. As the network is constrained to classify based on the aggregated vector, it is forced to extract the most salient features for each class. The attention blocks are generated from extracted features from different layers in the baseline network. Therefore, it preserves local information and spatial context or attended features in both coarse and fine scales. The network architecture's block diagram is presented in \textbf{Fig \ref{fig:AttentionBlockOption2}}. 
   
\begin{figure}[ht]
   \begin{center}
   \begin{tabular}{c} 
   \includegraphics[height=4cm]{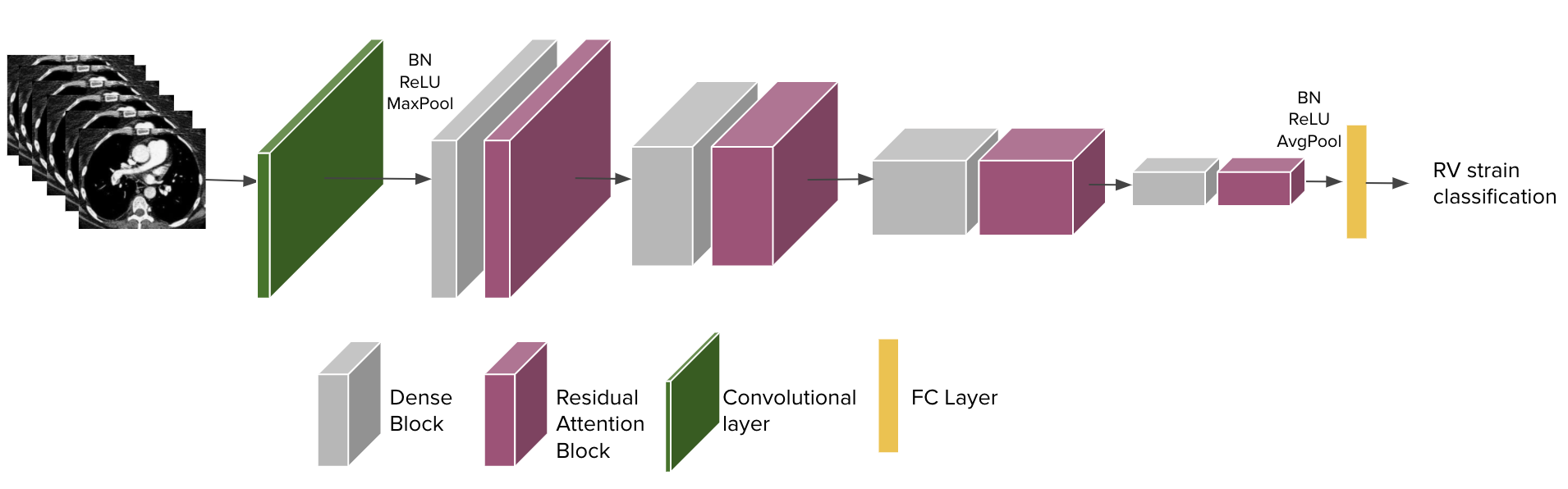}
   \end{tabular}
   \end{center}
   \caption[example] 
   { \label{fig:AttentionBlockOption1} 
SANet: 3D DenseNet with stacked attention model block diagram}
   \end{figure}

For the second network -- \textit{Stacked Attention Network} (SANet), we incorporate the residual attention blocks between the network’s layers in a stacked manner. \textbf{Fig ~\ref{fig:AttentionBlockOption1}} shows the block diagram of the developed system. In this approach, the blocks are added as an integrated part of the network and not as a separate branch of predictions, as in the previous architecture.
Due to this fact and the differentiability of the attention blocks, this network can be easily trained in an end-to-end manner. This network adaptively changes attention as the feature changes and layers go deeper. Each attention block learns and captures different types of features (will be presented in Sec.~\ref{sec:attention_contribution}). Thus, wrong attention predicted by one block can be masked by the other attention blocks. This makes a multiple block network of this sort quite robust to the attention prediction, as will be shown in Sec.~\ref{sec:results_multiple_attention}. The two approaches are compared in Sec.~\ref{sec:results_model}.


%
%

\subsection{3D Residual Attention Block}
The 3D attention block is comprised of two units: mask and trunk. The \textit{trunk branch} is constructed from ResNet residual block units 
to perform feature processing. The \textit{mask} is designed of soft attention structure\citep{NIPS2015_5854} that creates a 3D attention coefficients mask from the 3D features generated by the trunk. Thus, the mask behaves as a feature selector. The output of the attention module \(A(x)\) for input \(x\), which gives us the weighted feature structure, is created by the element-wise multiplication of trunk feature-maps \(T(x)\) and attention coefficients created by the mask layer \(M(x)\), as defined in equation \ref{eq:attention}:

\[ A_{i,d,c}(x) = M_{i,d,c}(x)*T_{i,d,c}(x)  \tag{1} \label{eq:attention}\]

Here:  \(i \in \{1, ..., H*W \} \) ranges over all spatial positions, \(H\) and \(W\) are the height and width of the scan, \(d \in \{1, ..., D \} \) is the slice index, and \(c \in \{1, ..., C \} \) is the index of the channel. The entire attention structure can be trained end-to-end. During  back-propagation, the attention mask serves as a gradient update filter due to its property of differentiability. In the soft mask branch, the gradient of the mask for the input features is shown in equation (\ref{eq:attention_derivative}). 

\[ \frac {\partial{M(\theta,x)T(\phi,x)}}{\partial \phi} = M(\theta,x)\frac{\partial T(\phi,x)}{\partial \phi} \tag{2} \label{eq:attention_derivative}\]

Where \(x\) is the input, \(\phi\) are trunk layer parameter, and \(\theta\) are mask layer parameters. The partial derivative of the trunk features \(T\) is multiplied by a factor of the mask \(M\). Therefore, if the trunk features are not correct the mask can prevent the features from updating the parameters. 

In our presented networks, we stack multiple attention blocks. This is designed for several reasons: First, RV strain classification from volumetric CTPA is a challenging task considering the number of features the network has to learn for a single scan as compared to a single image; The multiple attention network mitigates this problem by learning multiple masks. Second, the use of multiple attention blocks makes the network more robust. It can capture different attention types, focus on different features at each attention block, and correct a wrong prediction made from one block by other blocks.

For attention module stacking in SANet architecture, we need to use attention residual learning \citep{8100166}, rather than a naive multiplication as in equation (\ref{eq:attention}), as described in equation (\ref{eq:residual_attention}): 

\[ A_{i,d,c}(x) = (1 + M_{i,d,c}(x))*T_{i,d,c}(x)  \tag{3} \label{eq:residual_attention}\]

Here, similar to ideas in residual learning, adding ``1" to the generated mask \(M(x)\) preserves the identity function of the residual network, as the trunk \(T(x)\) is not multiplied with the zeros from the mask \(M(x)\). This allows us to stack and integrate multiple attention modules without the obvious performance drop of the straightforward multiplication. 

The mask is structured from a fully convolutional autoencoder design that resembles the popular U-Net \citep{Unet} used for image segmentation tasks. However, the difference between this structure and the U-Net is that our mask branch aims to improve trunk branch features rather than solve a complex problem directly. This structure creates a mask that acts as a filter to the trunk layer features. The mask branch contains an encoder which collects global information and a decoder part that combines global information with original feature maps. 3D max-pooling is applied several times for down-sampling after a small number of residual units. After reaching the lowest spatial resolution, the global information is then expanded by a symmetrical architecture. 3D interpolation layers up samples the output after some residual units. The number of 3D interpolation layers is the same as 3D max-pooling layers so that the output of the mask layer will have the same dimensions as the output from the trunk layer. Finally, a sigmoid layer normalizes the output range to [0, 1] after two consecutive 1 x 1 x 1 convolution layers. Skip connections are also added, connecting the encoder and decoder structures, to get the feature information from various scale levels. 

\begin{figure}[ht]
   \begin{center}
   \begin{tabular}{c} 
   \includegraphics[height=5cm]{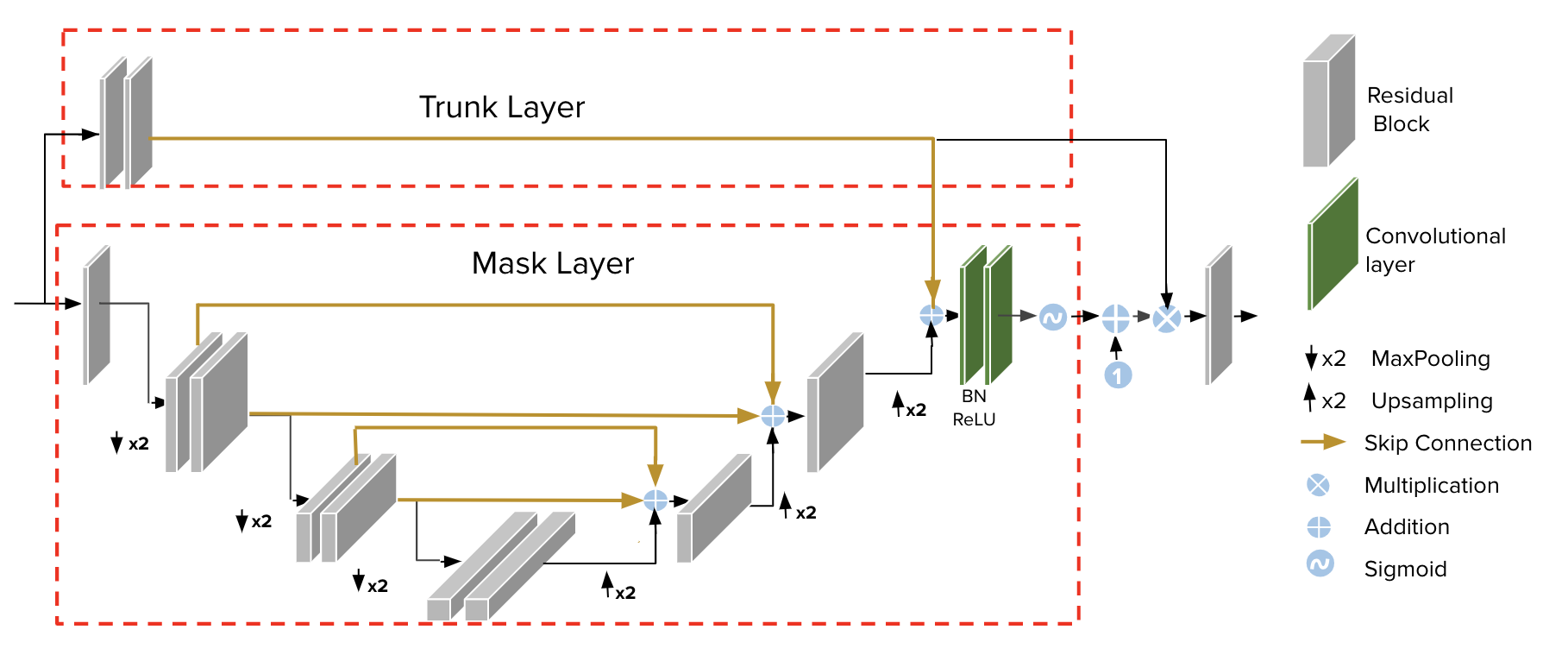}
   \end{tabular}
   \end{center}
   \caption[example] 
   { \label{fig:Attention} 
Our Proposed attention block architecture}
   \end{figure}

We add four attention blocks after each layer in the network. The number of downsampling (and corresponding upsampling) layers performed in each attention block is reduced as we go deeper into the network. The lowest spatial resolution is shared. For example, the attention block added after the first layer has four 3D max-pooling layers, whereas the block added after the second layer has only three layers, and so on. The full module for the second attention block is illustrated in \textbf{Fig ~\ref{fig:Attention}}.


\subsection{Dataset and Annotations}
We apply our suggested models to the specific challenging problem of RV strain classification from 3D CTPA scans. An institutional review board (IRB) approval was granted to this retrospective study. The IRB committee waived informed consent. We retrieved data of consecutive patients diagnosed with PE in our emergency department (ED). All the patients underwent CTPA between 1/2012 to 12/2018. All the scans were interpreted by board-certified radiologists. We used the CTPA scan report description to extract a single label of ``RV strain biomarker" and label the scans as either RV strain positive or RV strain negative. A scan was marked positive if the radiologist specifically indicated that RV strain was present. Selecting one high-level (or series-based) label for the whole 3D scan, with no additional markings or segmentation maps, enforces this computer-aided diagnosis (CAD) task to a weakly supervised solution and the processing of the scans in a 3D manner. Our dataset included 363 CTPAs, 86 of whom (23.91\%) were labeled with RV strain. The model was trained and validated on years 2012-2017 data and tested on held-out year 2018 data. Table ~\ref{tab:dataset} describes the dataset partitioning.

\begin{table}[ht]
\caption{Dataset specifications} 
\label{tab:dataset}
\begin{center}   
\scalebox{0.85}{
\begin{tabular}{|l||l|l|}
\hline
\rule[-1ex]{0pt}{3.5ex}  Group & Number of Images  & \% \\
\hline\hline
\rule[-1ex]{0pt}{3.5ex}  Total  (years 2012-2018) & 363 & 100 \\ 
\hline
\rule[-1ex]{0pt}{3.5ex}  Train  (years 2012-2017) &  248 & 68.3 \\ 
\hline
\rule[-1ex]{0pt}{3.5ex}  Validation  (years 2012-2017) & 44 & 12.2\\ 
\hline
\rule[-1ex]{0pt}{3.5ex}  Test (year 2018) & 71 & 19.5 \\
\hline 
\rule[-1ex]{0pt}{3.5ex}  Number of positives & 86 & 23.9 \\
\hline 
\end{tabular}}
\end{center}
\end{table}

\subsection{Evaluation methods}
The model's performance evaluation on the test set included AUC, sensitivity (also known as recall), specificity, accuracy, positive predictive value (PPV, also known as precision), and negative predictive value (NPV). The predicted probability threshold for returning a positive finding was determined by Youden's index \citep{youden}, which finds the model's optimal joined sensitivity and specificity. To measure the variability in these estimates, we calculated 95\% DeLong CIs \cite{delong} for the AUC of the model.

\subsection{Image Preprocessing}
Contrast-enhanced chest CT series were used as the dataset. Scans were extracted from Digital Imaging and Communications in Medicine (DICOM) format. Each extracted slice was scaled to 128 x 128 pixels. The entire series of 128 slices was saved as a 128 x 128 x 128 array. The scans are fed to the network in a 3D manner (as a whole), which takes advantage of the scan’s contextual context and allows for easy end-to-end training. An important task is to focus attention to a volume of interest (VOI) that includes the heart and lung regions to improve the developed system's performance. This enables us to choose a smaller volume to be fed into our network, which is especially important due to the computational costs of 3D processing. We chose the VOI according to the lung area. The lungs can be located in a chest CT image by using Hounsfield unit (HU) values. The heart area is roughly the area between the lungs. 

\subsection{Training and Inference}
To evaluate the compared networks fairly, we used the same experimental conditions. We used a batch size of 4 and a constant learning rate of 5e-3. Furthermore, we used Adam optimizer, weight decay of 1e-5, and binary cross-entropy loss \(L_{bce}\):

\[ L_{bce} = Y \log{Y_{pred}} + (1-Y)\log{(1-Y_{pred})} \tag{4} \label{eq:loss}\]

Where \(Y_{pred} \in (0.0, 1.0) \) represents the the predicted probability, and \(Y \in (0,1) \) is the binary ground truth label. We used data augmentations of flip and rotation. Our code is implemented in Pytorch. The network was trained on an NVIDIA 1080Ti GPU.

\section{EXPERIMENTS AND RESULTS}
\label{sec:results}

\subsection{Model Performance}
\label{sec:results_model}

We evaluated the performance of SANet and MLANet. The comparison between the two model architectures is described in Table ~\ref{tab:network_comparison}. As can be seen from the results, adding our self-attention blocks improves performance compared to both approaches' baseline network. In addition, the stacked attention block solution - SANet, outperforms MLANet for all baseline networks. For both models, the DenseNet121-3D baseline attains the best results. SANet with the baseline of 3D-DenseNet-121 achieved the best performance, with an AUC of 0.88 (95\% CI: 0.8-0.94) for classifying RV strain. 84.7\% of the test cases are correctly predicted according to the report label. Using Youden's index, the model showed a sensitivity of 87\% and specificity of 83.7\% for predicting RV strain. The solution's precision was 71.4\%, and NPV was 93.2\%. 
The SANet DenseNet121-3D baseline outperforms the next best SANet model with a margin of 0.045 AUC and the MLANet DenseNet backbone with a gain of 0.064 AUC.

   \begin{table}[ht]
\caption{Comparison between SANet and MLANet performance with 95\% confidence interval } 
\label{tab:network_comparison}
\begin{center}   
\scalebox{0.85}{
\begin{tabular}{|l||l|l|l|l|l|l|}
\hline
\rule[-1ex]{0pt}{3.5ex}  Backbone & AUC & Accuracy & Specificity & Sensitivity & PPV & NPV  \\
\hline\hline
\multicolumn{7}{|c|}{SANet}\\\hline
\hline
\rule[-1ex]{0pt}{3.5ex}  ResNeXt3D-101 & $0.788$ [$0.69$-$0.88$] & $0.68$ & $0.592$ & $0.869$ & $0.5$ & $0.906$ \\ 
\hline
\rule[-1ex]{0pt}{3.5ex}  DenseNet3D-121 & $\textbf{0.877 [0.81-0.95]}$ & $\textbf{0.847}$ & $\textbf{0.837}$ & $\textbf{0.87}$ & $\textbf{0.714}$ & $\textbf{0.932}$\\
\hline
\rule[-1ex]{0pt}{3.5ex}  ResNet3D-50 (basic block)  & $0.803$ [$0.71$-$0.9$] & $0.75$ & $0.735$ & $0.783$ & $0.581$ & $0.878$ \\ 
\hline
\rule[-1ex]{0pt}{3.5ex}  ResNet3D-50 (residual block) & $0.832$ [$0.75$-$0.91$] & $0.778$ & $0.735$ & $0.87$ & $0.606$ & $0.923$ \\ 
\hline\hline
\multicolumn{7}{|c|}{MLANet}\\\hline
\hline
\rule[-1ex]{0pt}{3.5ex}  ResNeXt3D-101 & $0.76$ [$0.66$-$0.86$] & $0.75$ & $0.714$ & $0.826$ & $0.576$ & $0.897$\\ 
\hline
\rule[-1ex]{0pt}{3.5ex}  DenseNet3D-121 & $\textbf{0.813 [0.71-0.9]}$ & $\textbf{0.764}$ & $\textbf{0.735}$ & $\textbf{0.826}$ & $\textbf{0.594}$ & $\textbf{0.9}$ \\ 
\hline
\rule[-1ex]{0pt}{3.5ex}  ResNet3D-50 (basic block) & $0.782$ [$0.68$-$0.87$] & $0.694$ & $0.633$ & $0.826$ & $0.514$ & $0.633$\\ 
\hline
\rule[-1ex]{0pt}{3.5ex}  ResNet3D-50 (residual block) & $0.777$ [$0.66$-$0.88$] & $0.75$ & $0.735$ & $0.783$ & $0.581$ & $0.878$ \\ 
\hline 
\end{tabular}}
\end{center}
\end{table}

\subsection{Interpretation of the Model Prediction}
\begin{figure}[ht]
   \begin{center}
   \begin{tabular}{c} 
   \includegraphics[height=7cm]{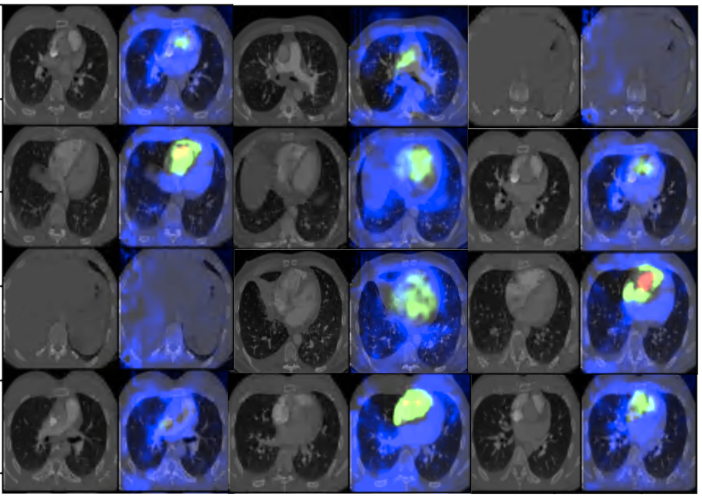}
   \end{tabular}
   \end{center}
   \caption[example] 
   { \label{fig:grad_cam2} 
Heatmap visualization results of correctly classified examples: Pairs of 2D slices taken from 3D CTA scans and their corresponding heatmap results. High activations (red, yellow and green) indicate the areas on which our trained CNN picked up as main features for classification.  The presented results were obtained from aggregating features from all layers and combining them via multiplication and normalization.}
   \end{figure}

In order to obtain model interpretability, we identified locations and slices in the scan that contributed most to the classification using 3D gradient-weighted class activation mapping (Grad-CAMs) \citep{Grad_Cam}. We expanded the implementation from a 2D slice-based tool to a 3D one. \textbf{Fig ~\ref{fig:grad_cam2}} shows the heatmap visualization results obtained from our network. The brightest areas of the heatmap are regions that influence the model prediction the most. We note that the attention gate focuses on the heart area rather than other scan regions, as expected. 


\subsection{Attention Contribution}
\label{sec:attention_contribution}
Table \ref{tab:attention_results} examines the result of adding our proposed attention blocks to common baseline architectures in a stacked approach. As shown in the table, adding our proposed attention blocks consistently improves performance with a maximum gain of 0.075 in AUC. All improvements are statistically significant except for the ResNet3D-50 (residual block) baseline. Our proposed solution is constructed using a backbone of densely connected blocks, as in the DenseNet3D-121 baseline, which achieves the best results. It outperforms all other models with a margin of 0.045 AUC.


\begin{table}[ht]
\caption{Comparing the AUC score (with 95\% confidence interval) effect of adding stacked residual attention modules (SANet) to different baseline architectures}
\label{tab:attention_results}
\begin{center}
\scalebox{0.85}{
\begin{tabular}{|l||l|l|l|}
\hline
\rule[-1ex]{0pt}{3.5ex}  Network & baseline & with attention & P-value  \\
\hline\hline
\rule[-1ex]{0pt}{3.5ex}  ResNext3D-101 \citep{Xie2017AggregatedRT} & $0.713$ [$0.61$-$0.81$] & $0.788$ [$0.69$-$0.88$] & \it{p} \textless{ $0.001$}\\ 
\hline
\rule[-1ex]{0pt}{3.5ex}  DenseNet3D-121 \citep{huang2018densely} & $0.805$ [$0.7$-$0.89$] & $\textbf{0.877 [0.81-0.95]}$  & \it{p} \textless{ $0.005$}\\ 
\hline
\rule[-1ex]{0pt}{3.5ex}  ResNet3D-50 (basic block) \citep{DBLP:journals/corr/HeZRS15} &  $0.765$ [$0.66$-$0.87$] & $0.803$ [$0.71$-$0.9$] &  \it{p} \textless{ $0.01$}\\ 
\hline
\rule[-1ex]{0pt}{3.5ex}  ResNet3D-50 (residual block) & $0.809$ [$0.7$-$0.89$] & $0.832$  [$0.75$-$0.91$] & \it{p not significant}\\ 
\hline
\end{tabular}}
\end{center}
\end{table}

\begin{figure}[ht]
   \begin{center}
   \begin{tabular}{c} 
   \includegraphics[height=7cm]{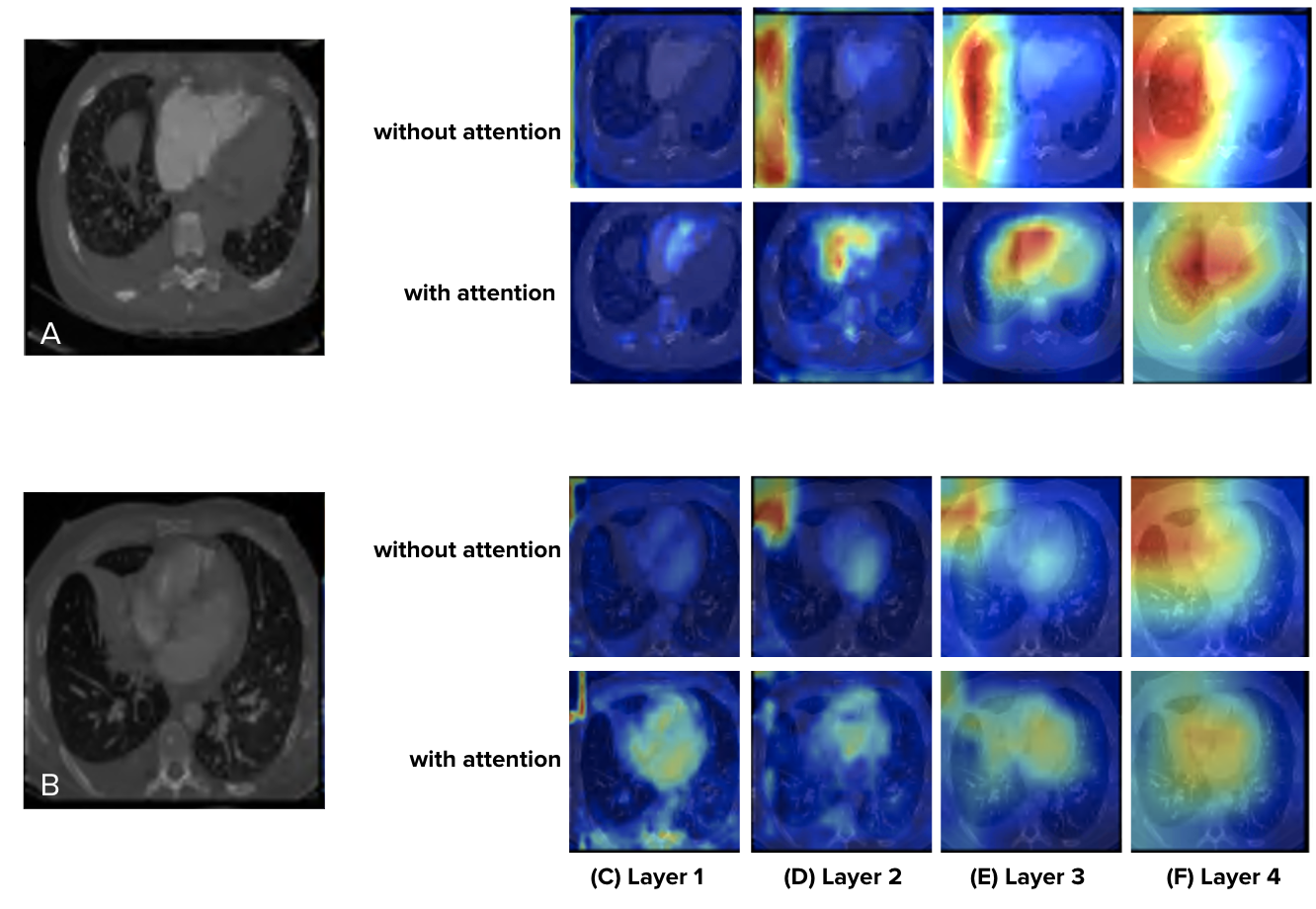}
   \end{tabular}
   \end{center}
   \caption[example] 
   { \label{fig:grad_cam_compare2} 
The visual comparisons of scan slices Grad-Cam visualization examples of multi-level maps from DenseNet3D-121 (rows 1 and 3) and proposed attentive maps from SANet (rows 2 and 4). (A) and (B) are two CTA scan slice images; (C)-(F) show the output Grad-Cam maps from layer 1 (shallow layer) to layer 4 (deep layer) of the networks. We can observe that directly applying multi-level features without attention blocks may suffer from poor localization. In contrast, our proposed attentive maps are more powerful for the better representation of the heart area in the scan.}
   \end{figure}

\textbf{Fig ~\ref{fig:grad_cam_compare2}} visualizes the comparison between DenseNet3D-121 and SANet. The figure visualizes the Grad-Cam maps extracted from different layers. The areas on which the networks center on different layers can be seen, as well as the attention blocks contribution as it shifts the network's focus to the heart area.


\subsection{Comparison with state-of-the-art 3D CNN models}
We compare SANet to several common 3D CNN architectures as detailed in Table \ref{tab:3D_CNN_results}. This includes the current state-of-the-art model architecture for Kinetics-600 dataset \citep{carreira2018short}, ResNeXt3D-101, and memory-efficient DenseNet3D-121, and the classic ResNet3D-50 in both its basic and residual block version. 
In addition, we compare our network to two promising recently published 3D classification networks as well: Med3D \citep{DBLP:journals/corr/abs-1904-00625}, a ResNet3D-50 network-based that is pretrained using many medical image datasets and uses transfer learning to create a robust network for medical domain 3D classification tasks and Models Genesis \citep{zhou2019models}, a U-Net architecture based solution that is pretrained using a pipeline of self-supervision tasks.
Our model outperforms all compared models with a wide margin of 0.068 in AUC. P-Value with compare to SANet shows that all results are statistically significant.

   \begin{table}[ht]
\caption{Comparison with state-of-the-art 3D CNN models} 
\label{tab:3D_CNN_results}
\begin{center}   
\scalebox{0.85}{
\begin{tabular}{|l||l|l|l|l|l|l|l|}
\hline
\rule[-1ex]{0pt}{3.5ex}  Model Architecture & AUC & Accuracy & Specificity & Sensitivity & PPV & NPV  & P-Value\\
\hline\hline
\rule[-1ex]{0pt}{3.5ex}  ResNet3D-50 (basic block) & $0.765$ [$0.66$-$0.87$] & $0.722$ & $0.694$ & $0.783$ & $0.545$ & $0.872$ & \it{p} \textless{ $0.005$}\\ 
\hline
\rule[-1ex]{0pt}{3.5ex}  ResNet3D-50 (residual block) & $0.809$ [$0.7$-$0.89$] & $0.778$ & $0.837$ & $0.652$ & $0.652$ & $0.837$ & \it{p} \textless{ $0.05$}\\
\hline
\rule[-1ex]{0pt}{3.5ex}  ResNeXt3D-101 & $0.713$ [$0.61$-$0.81$] & $0.681$ & $0.673$ & $0.696$ & $0.5$ & $0.825$ & \it{p} \textless{ $0.001$}\\
\hline
\rule[-1ex]{0pt}{3.5ex}  DenseNet3D-121 & $0.805$ [$0.7$-$0.89$] & $0.792$ & $0.796$ & $0.783$ & $0.643$ & $0.886$ & \it{p} \textless{ $0.05$}\\
\hline 
\rule[-1ex]{0pt}{3.5ex}  Models Genesis & $0.796$ [$0.69$-$0.89$] & $0.681$ & $0.592$ & $0.87$ & $0.5$ & $0.906$ & \it{p} \textless{ $0.03$}\\
\hline 
\rule[-1ex]{0pt}{3.5ex}  Med3D & $0.801$ [$0.7$-$0.89$] & $0.75$ & $0.714$ & $0.826$ & $0.576$ & $0.897$ & \it{p} \textless{ $0.05$}\\
\hline 
\rule[-1ex]{0pt}{3.5ex}  SANet & $\textbf{0.877 [0.81-0.95]}$ & $\textbf{0.847}$ & $\textbf{0.837}$ & $\textbf{0.87}$ & $\textbf{0.714}$ & $\textbf{0.932}$ & --\\
\hline 
\end{tabular}}
\end{center}
\end{table}

\begin{figure}[ht]
   \begin{center}
   \begin{tabular}{c} 
   \includegraphics[height=7cm]{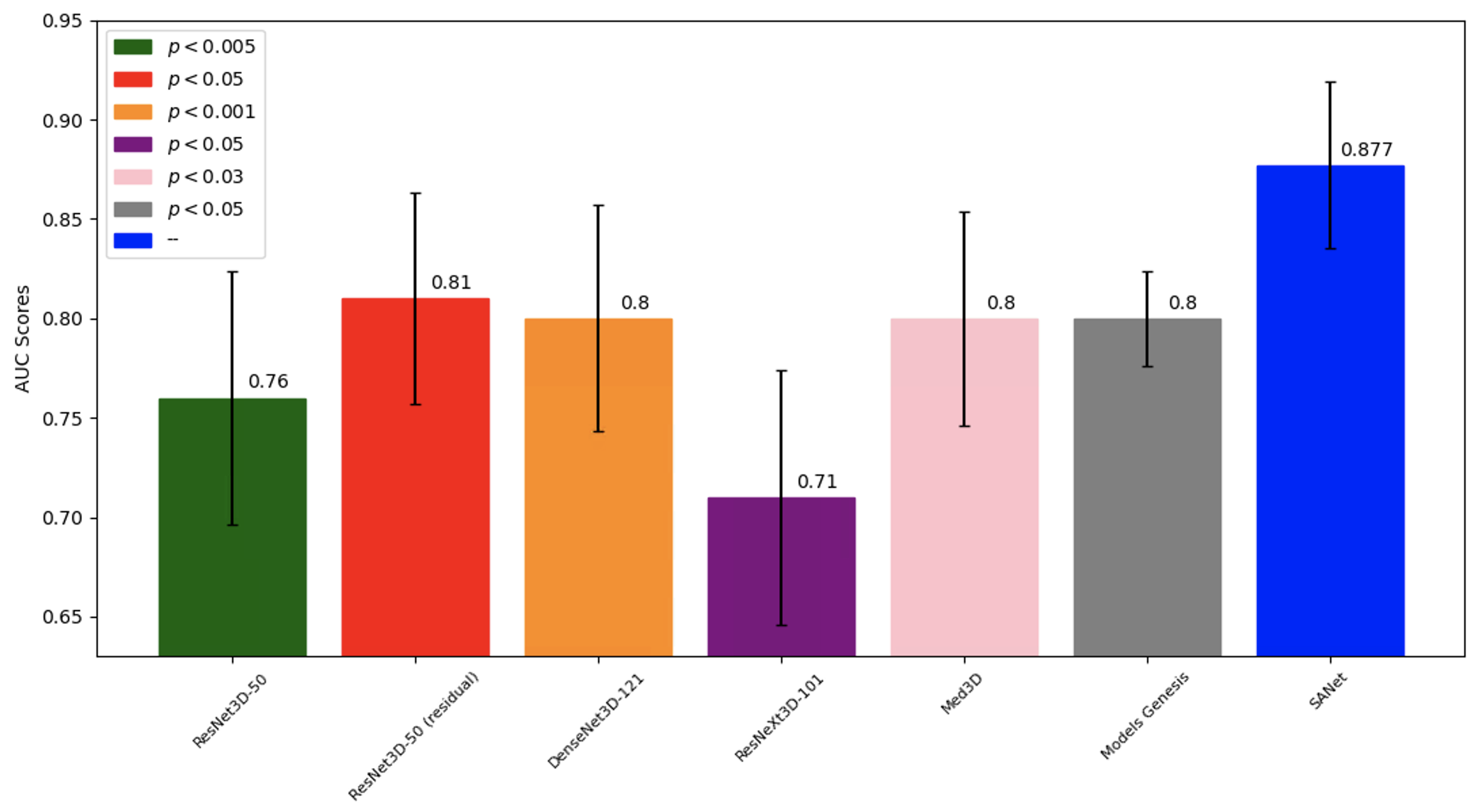}
   \end{tabular}
   \end{center}
   \caption[example] 
   { \label{fig:auc_compare2} 
ROC comparison with state-of-the-art 3D CNN models}
   \end{figure}

In \textbf{Fig ~\ref{fig:auc_compare2}} the different AUC scores are displayed, allowing visual comparison. 


\subsection{Ablation Study On Multiple Attention Blocks}
\label{sec:results_multiple_attention}
To further analyze the system, we explore altering the number of attention blocks fused into our model. This enables a comparison to the baseline network version which has the same structure but has zero attention blocks. Table ~\ref{tab:attention_number_results} presents the results of this experiment. The entries in the table specify the best performance for each number of added attention blocks, where the number in the brackets list the exact blocks that where added: Adding the described attention blocks shows improvement in result metrics. It is evident that four attention blocks perform better than the networks having one or two blocks. This result follows intuition and can be explained by the fact that multiple attention blocks tend to obtain the correct attention and can even compensate over the wrong feature capture one of the attention blocks. Although networks with three and four blocks attain almost the same AUC, all other metrics are improved when adding a fourth block. In addition, when integrating an attention block after the first layer improves the AUC most. Looking at the results, it is only expected that we add more blocks until saturation or degradation in performance. Yet, we are limited by our baseline's depth, DenseNet3D-121, which includes only four basic units.

   \begin{table}[ht]
\caption{Residual attention model comparison with different number of stacked attention blocks (SANet)} 
\label{tab:attention_number_results}
\begin{center}   
\scalebox{0.85}{
\begin{tabular}{|l||l|l|l|l|l|l|}
\hline
\rule[-1ex]{0pt}{3.5ex}  SANet Architecture & AUC & Accuracy & Specificity & Sensitivity & PPV & NPV  \\
\hline\hline
\rule[-1ex]{0pt}{3.5ex}  DenseNet121-3D & $0.805$ & $0.792$ & $0.796$ & $0.783$ & $0.643$ & $0.886$  \\
\hline
\rule[-1ex]{0pt}{3.5ex}  DenseNet121-3D with 1 Attention Blocks [1] & $0.831$ & $0.792$ & $0.816$ & $0.739$ & $0.654$ & $0.87$ \\
\hline
\rule[-1ex]{0pt}{3.5ex}  DenseNet121-3D with 2 Attention Blocks [1,2] & $0.86$ & $0.778$ & $0.735$ & $0.87$ & $0.606$ & $0.923$ \\
\hline
\rule[-1ex]{0pt}{3.5ex}  DenseNet121-3D with 3 Attention Blocks [1,2,4] & $0.868$ & $0.833$ & $0.816$ & $0.87$ & $0.69$ & $0.93$ \\
\hline
\rule[-1ex]{0pt}{3.5ex}  SANet & $\textbf{0.877}$ & $\textbf{0.847}$ & $\textbf{0.837}$ & $\textbf{0.87}$ & $\textbf{0.714}$ & $\textbf{0.932}$\\
\hline
\end{tabular}}
\begin{tablenotes}
\item [1] The numbers in the brackets indicate the attention blocks that were added to the DenseNet3D-121 baseline. For example: [1,3] means that two attention blocks were added after the first and third layers. 
\end{tablenotes}
\end{center}
\end{table}

\section{CONCLUSIONS}

In this work, we proposed generalized self-attention blocks that can be incorporated into existing classification architectures. The use of residual networks allows increasing the network's depth while keeping it stable for training, improving performance significantly \citep{DBLP:journals/corr/HeZRS15}. Dense connectivity ensures maximum information paths between layers by connecting all layers \citep{huang2018densely}. The output features of all convolution layers in the dense blocks are concatenated along the channel axis. Attention blocks automatically learn to focus on target structures without additional supervision. They do not introduce significant computational overhead and do not require a large number of model parameters, as in the case of multi-model or cascaded frameworks. In return, the proposed attention blocks improve the model's sensitivity and accuracy for label predictions by suppressing feature activations in irrelevant regions. In this way, the necessity of using an additional external organ localization model can be eliminated while maintaining the high prediction accuracy. All of these motivated us to use these mechanisms for our proposed network.

We present two network variations that utilize our proposed blocks to classify RV strain on CTPA scans: MLANet - The multi-layer attention block architecture in which the attention blocks are aggregated from different layers of the network for prediction and SANet - the stacked attention block architecture where the attention block is incorporated into the network as additional network layers. Our results show that attention consistently improves performance throughout multiple baseline architectures. The stacked attention block model with a baseline of 3D-DenseNet-121 outperforms the multi-layer architecture with an additional AUC margin of 0.064. This network also outperforms current state-of-the-art 3D CNN models: ResNeXt3D-101, ResNet3D-50 and DenseNet3D-121 for the same task. We conclude that the automatic deep learning solution can classify RV strain effectively from 3D CTPA scans.

We selected a single binary label from the radiologist report. This means that readings were done in an acute scenario rather than a prospective research environment. On the positive side, if our system was able to cope with the provided settings it may provide a more robust solution that can be translatable in real-world settings.



Beyond RV strain's specific task, this approach is also applicable to other abnormality classification problems with minimal preprocessing and a single binary scan-level label of the 3D volumetric data. We plan to test these generalization abilities in our future work and specifically use the system proposed here for the broader PE risk stratification task. Such tools may support improved healthcare. For example, improving the ability to direct preventative and health surveillance resources. To the best of our knowledge, there is no prior deep learning-based solution for fully automated classification of RV strain using CTPA, and this is the first work where medical images are used in such an architecture.

%

\acknowledgments 
The research in this publication was supported by the Israeli Science Foundation (ISF) under Grant number 20/2629.

\bibliography{report} 
\bibliographystyle{spiebib} 

\end{document}